\documentclass{PoS}

\title{Non-zero density QCD by the Taylor expansion method: The isentropic equation of state, hadronic fluctuations and more}

\ShortTitle{Non-zero density QCD by the Taylor expansion method}

\author{\speaker{\bf Chuan Miao and Christian Schmidt} ({\bf RBC-Bielefeld Collaboration})\\
	Department of Physics, Brookhaven National Laboratories, Upton, NY 11733, USA.\\
        Fakult\"at f\"ur Physik, Universit\"at Bielefeld, D-33615 Bielefeld, Germany.\\
	E-mail: \email{chuan@quark.phy.bnl.gov}, \email{schmidt@physik.uni-bielefeld.de}}

\abstract{
We discuss the Taylor expansion approach to non-zero baryon chemical
potential ($\mu_B$) and present results on expanion coefficients of
the pressure and energy density up to the 6-th order in
$\mu_B$. Calculations have been performed with (2+1)-flavor of
improved staggerd fermions (p4fat3) and almost physical masses on
lattices with temporal extent of four and six time slices. We use
the expansion coefficients to construct the isentropic equation of state on lines of
constant entropy per baryon number and various different hadronic
fluctuations.  Furthermore, we estimate the
radius of convegence of our expansion, which can be seen as a method
to analyze the structure of the QCD phase diagram.  }

\FullConference{The XXVI International Symposium on Lattice Field Theory\\
		 July 14-19 2008\\
		 Williamsburg, Virginia, USA}

\usepackage{amsmath}

\newcommand{\muBQS}{\mu_{B,Q,S}}
\def\lsim{\raise0.3ex\hbox{$<$\kern-0.75em\raise-1.1ex\hbox{$\sim$}}}
\def\gsim{\raise0.3ex\hbox{$>$\kern-0.75em\raise-1.1ex\hbox{$\sim$}}}
\def\<{\left<}
\def\>{\right>}

\def\muBQS{\mu_{B,Q,S}}

\begin{document}
\section{Introduction}
A detailed and comprehensive understanding of the thermodynamics
of quarks and gluons, e.g. of the equation of state is most desirable
and of particular importance for the phenomenology of relativistic
heavy ion collisions.
Lattice regularized QCD simulations at non-zero temperatures have
been shown to be a very successful tool in analyzing the
non-perturbative features of the quark-gluon plasma. Driven by both,
the exponential growth of the computational power of recent
super-computer as well as by drastic algorithmic improvements one is
now able to simulate dynamical quarks and gluons on fine lattices with
almost physical masses. 

At non-zero chemical potential, lattice QCD is harmed by the
``sign-problem'', which makes direct lattice calculations with
standard Monte Carlo techniques at non-zero density practically
impossible.  However, for small values of the chemical potential, some
methods have been successfully used to extract information on the
dependence of thermodynamic quantities on the chemical potential.
For a recent overview see, e.g. \cite{overview}.

\section{The Taylor expansion method}
We closely follow here the approach and notation used in
Ref.~\cite{eos6}. We start with a Taylor expansion for the pressure in
terms of the quark chemical potentials
\begin{equation}
\frac{p}{T^{4}}
=\sum_{i,j,k}c^{u,d,s}_{i,j,k}(T)\left(\frac{\mu_{u}}{T}\right)^{i}
\left(\frac{\mu_{d}}{T}\right)^{j}\left(\frac{\mu_{s}}{T}\right)^{k}.
\label{eq:PTaylor}
\end{equation}
The expansion coefficients $c^{u,d,s}_{i,j,k}(T)$ are computed on the
lattice at zero chemical potential, using stochastic estimators. Some
details on the computation are given in~\cite{details1,details2}. 
Details on our current
data set and the number of random vectors used for
the stochastic random noise method are summarized in
Table~\ref{tab:stat}.
\begin{table}[!tbh]
\begin{center}
\begin{tabular}{|c|c|c|c||c|c|c|c|}
\hline
\multicolumn{4}{|c||}{$N_\tau=4$} & \multicolumn{4}{|c|}{$N_\tau=6$}  \\
\hline
T[MeV] & \#Conf. & Sep. & \#r.v. & T[MeV] & \#Conf. & Sep. & \#r.v. \\
\hline \hline
176.04 & 1013 & 20 & 480 & 173.82 & 985  & 10 & 400 \\
186.41 & 1550 & 30 & 480 & 179.63 & 910  & 10 & 400 \\
190.83 & 1550 & 30 & 480 & 185.64 & 1043 & 10 & 400 \\
195.37 & 1550 & 30 & 384 & 194.97 & 924  & 10 & 400 \\
202.42 & 1350 & 30 & 384 & 201.35 & 873  & 10 & 350 \\
204.83 & 475  & 60 & 384 & 204.58 & 717  & 10 & 200 \\
209.63 & 264  & 60 & 384 & 211.11 & 690  & 10 & 150 \\
218.18 & 365  & 30 & 384 & 224.34 & 560  & 10 & 150 \\
260.74 & 199  & 60 & 192 & 237.72 & 670  & 10 & 100 \\
306.88 & 302  & 60 & 96  & 278.36 & 540  & 10 & 50  \\
428.66 & 618  & 10 & 48  & 362.87 & 350  & 10 & 50  \\
       &      &    &     & 415.83 & 345  & 10 & 50  \\ 
\hline
\end{tabular}
\end{center}
\caption{Details on the calculation: The columns give from left to
right the temperature values, the number of
evaluated configurations, the number of trajectories by which these
configurations are separated and the number of random vectors used for
the evaluation of the traces, for $N_\tau=4$ and $6$, respectively. }
\label{tab:stat}
\end{table}

\begin{figure}
\begin{center}
\resizebox{0.32\textwidth}{!}{%
  \includegraphics{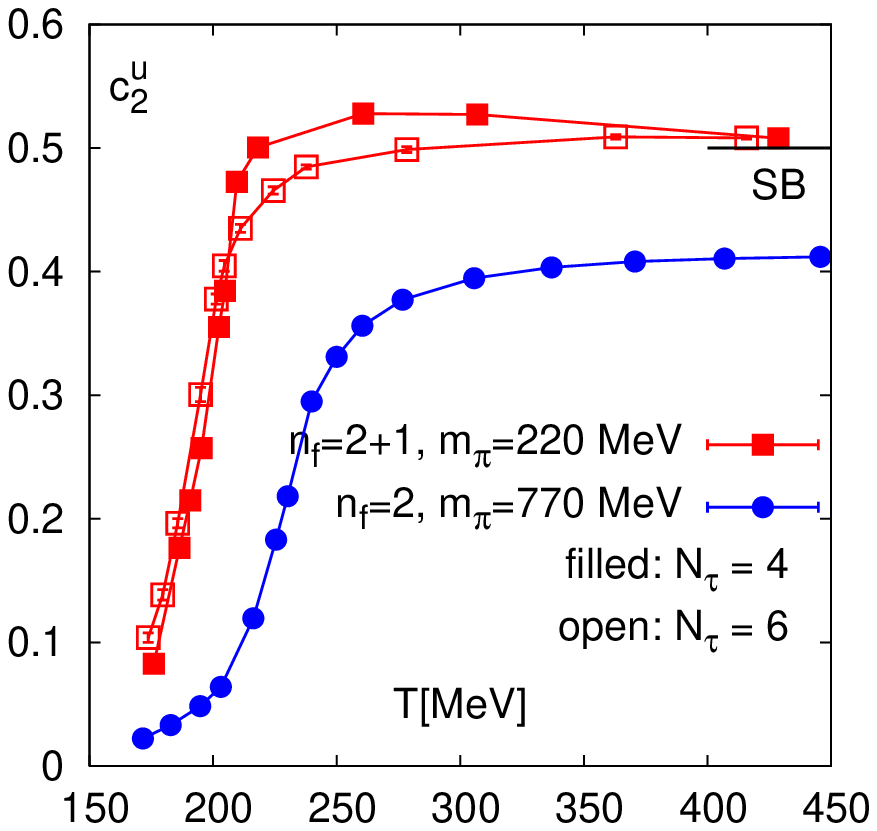}
}
\resizebox{0.32\textwidth}{!}{%
  \includegraphics{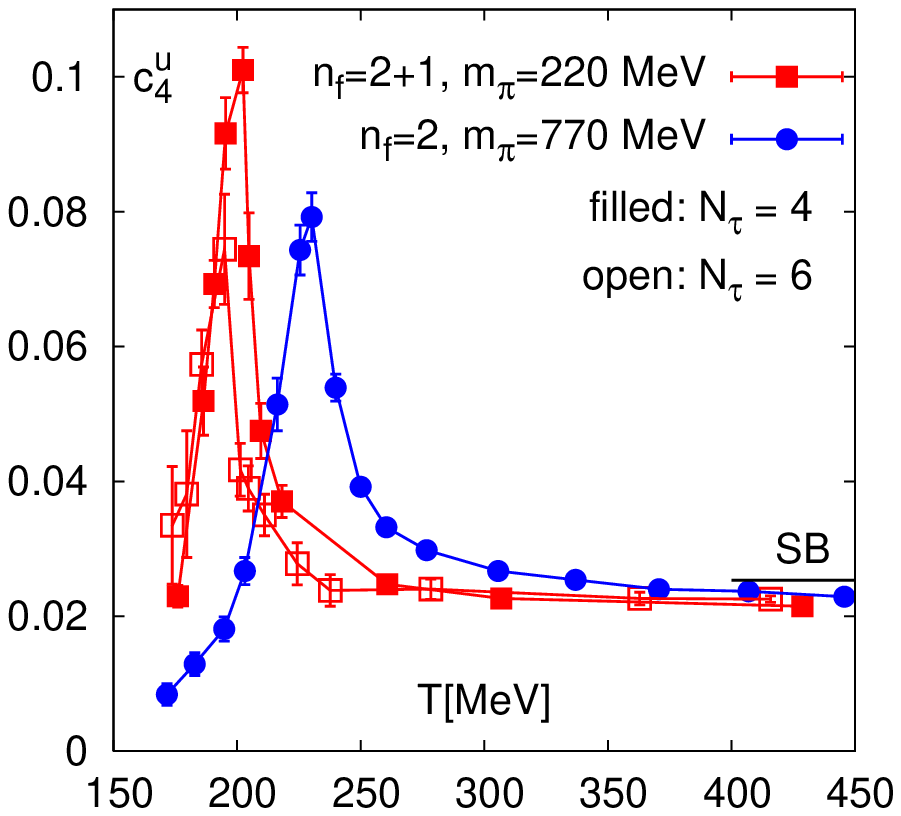}
}
\resizebox{0.32\textwidth}{!}{%
  \includegraphics{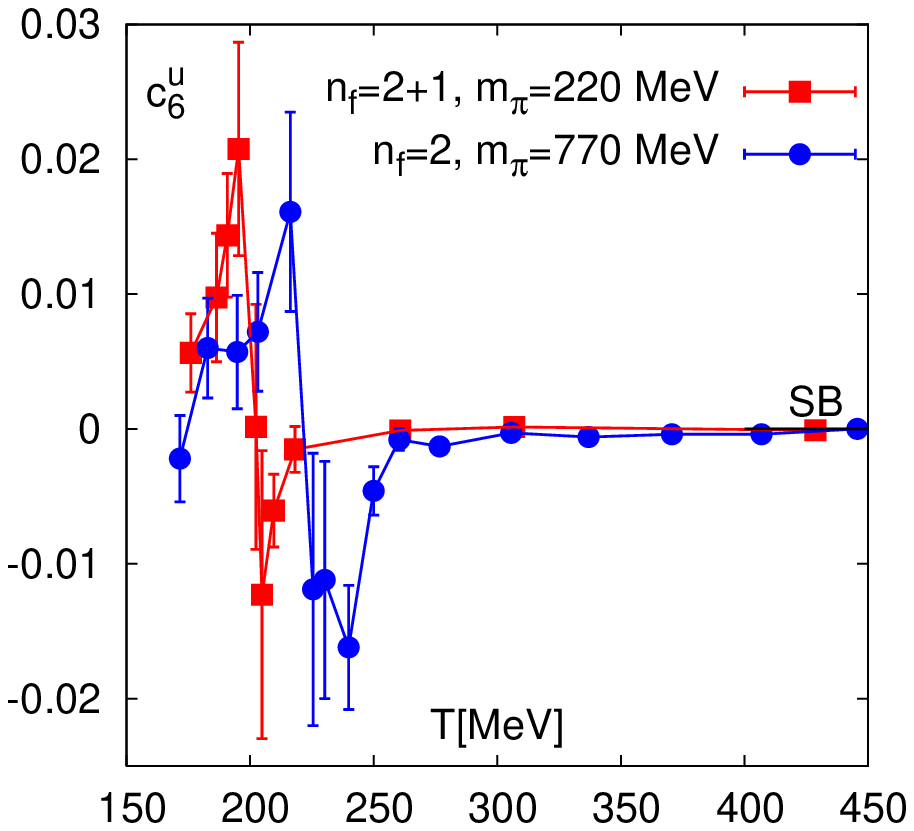}
}
\end{center}
\caption{Taylor coefficients of the pressure in term of the up-quark
chemical potential. Results are obtained with the p4fat3 action on
$N_\tau=4$ (full) and $N_\tau=6$ (open symbols) lattices. We compare
preliminary results of (2+1)-flavor a pion mass of $m_\pi\approx
220$~MeV to previous results of 2-flavor simulations with a
corresponding pion mass of $m_p\approx770$ \cite{eos6}.}
\label{fig:coff_uds}  
\end{figure}

In Fig.~\ref{fig:coff_uds} 
we show results on the diagonal expansion coefficients with respect to
the up-quark chemical potential up to the six order
($c_{n,0,0}^{u,d,s}$ with $n=2,4,6$). They have been evaluated on
configurations generated for a calculation of the equation of
state~\cite{EoS}, obtained with (2+1)-flavor of the p4fat3 action
\cite{p4fat3} and a pion mass of $m_\pi\approx 220$~MeV.  Here the
full symbols are from $N_\tau=4$ lattices, while the open symbols
denote results from $N_\tau=6$ lattices.  The spatial lattice size has
been chosen to be $N_\sigma=4N_\tau$, which is known to be large enough
in order to suppress  finite volume effects.  Considering cut-off
effects, we find that they are small and of similar magnitude as those
found for the trace anomaly \cite{EoS}. This was already anticipated
by the analysis of the cut-off corrections that arises in the free
gas limit \cite{Hegde}.  Similar results for the expansion
coefficients have been also obtained with the asqtad~\cite{milc_dens}
and the standard action~\cite{GG}.

We also compare our preliminary results for (2+1)-flavor QCD and a pion
mass of $m_\pi\approx 220$~MeV, with previously obtained result of
2-flavor QCD and $m_\pi\approx 770$~MeV \cite{eos6} (also p4fat3 and $N_\tau=4$).
It is apparent from
Fig.~\ref{fig:coff_uds} that the critical temperature for these
two particular sets of lattice parameter differ substantially.
In fact, the transition temperature decreases from about $225$~MeV for 
the heavier mass calculations
to about $200$~MeV for the lighter mass calculations. Note, that those
$T_c$ values are the $N_\tau=4$ values, which of course are still influenced by the 
finite lattice spacing.  Furthermore, we find from
Fig.~\ref{fig:coff_uds} that the quark number fluctuations of second,
fourth and sixth order, which are related to these expansion
coefficients, increase with decreasing quark mass.

Alternatively to the quark chemical potentials one can introduce
chemical potentials for the conserved quantities baryon number $B$,
electric charge $Q$ and strangeness $S$ ($\mu_{B,Q,S}$), which are
related to $\mu_{u,d,s}$ via
\begin{equation}
\mu_{u}=\frac{1}{3}\mu_{B} +\frac{2}{3} \mu_{Q},\qquad
\mu_{d}=\frac{1}{3}\mu_{B}-\frac{1}{3}\mu_{Q},\qquad
\mu_{s}=\frac{1}{3}\mu_{B}-\frac{1}{3}\mu_{Q}-\mu_{S}.
\label{eq:chempot}
\end{equation}
By means of these relations the coefficients $c^{B,Q,S}_{i,j,k}$ of
the pressure expansion in terms of $\mu_{B,Q,S}$ are easily obtained,
in analogy to Eq.~\ref{eq:PTaylor}
\begin{equation}
\frac{p}{T^{4}}
=\sum_{i,j,k}c^{B,Q,S}_{i,j,k}(T)\left(\frac{\mu_{B}}{T}\right)^{i}
\left(\frac{\mu_{Q}}{T}\right)^{j}\left(\frac{\mu_{S}}{T}\right)^{k}.
\label{eq:PTaylor_hadronic}
\end{equation}
For the rest of this article we will restrict ourselves to the case of 
$\mu_Q\equiv\mu_S\equiv0$, thus we will suppress in the following the 
indices that are related to these chemical potentials. From the pressure 
we immediately obtain the baryon number density
$n_B$, which is given by the derivative of $p/T^4$ with respect to the
baryon chemical potential $\mu_B$ and can be expressed in term of the
expansion coefficients $c_{n}^{B}$, we have
\begin{equation}
\frac{n_B}{T^{3}}
=\sum_{n=2}^{\infty}nc^{B}_{n}(T)\left(\frac{\mu_{B}}{T}\right)^{n-1}.
\label{eq:nB_hadronic}
\end{equation}

\begin{figure}
\begin{center}
\resizebox{0.32\textwidth}{!}{%
  \includegraphics{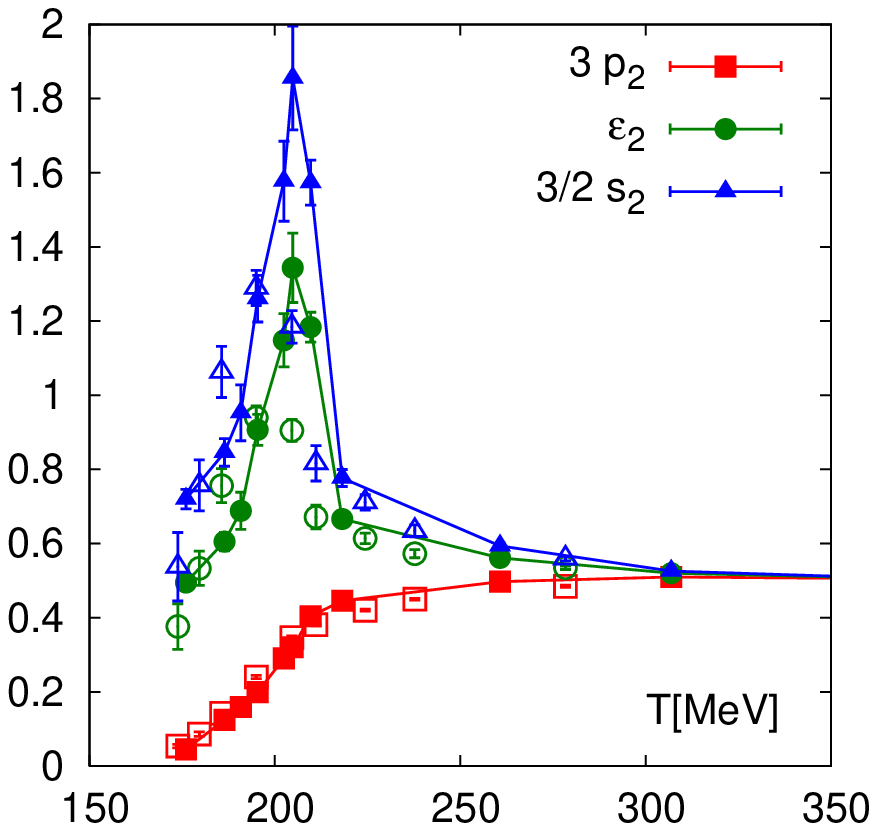}
}
\resizebox{0.32\textwidth}{!}{%
  \includegraphics{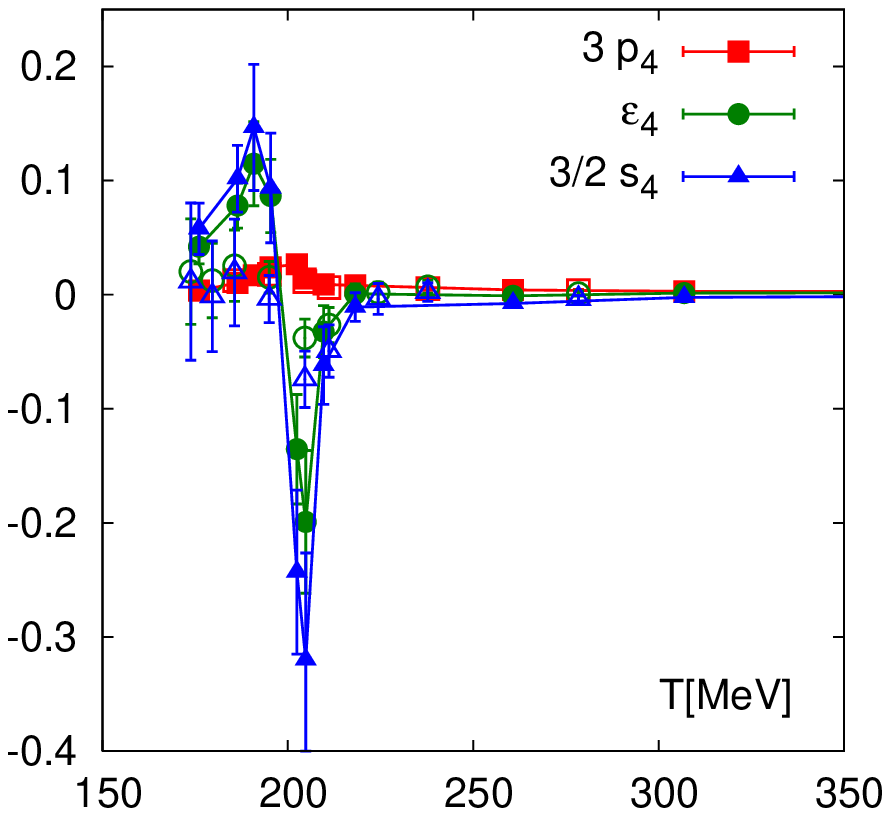}
}
\resizebox{0.32\textwidth}{!}{%
  \includegraphics{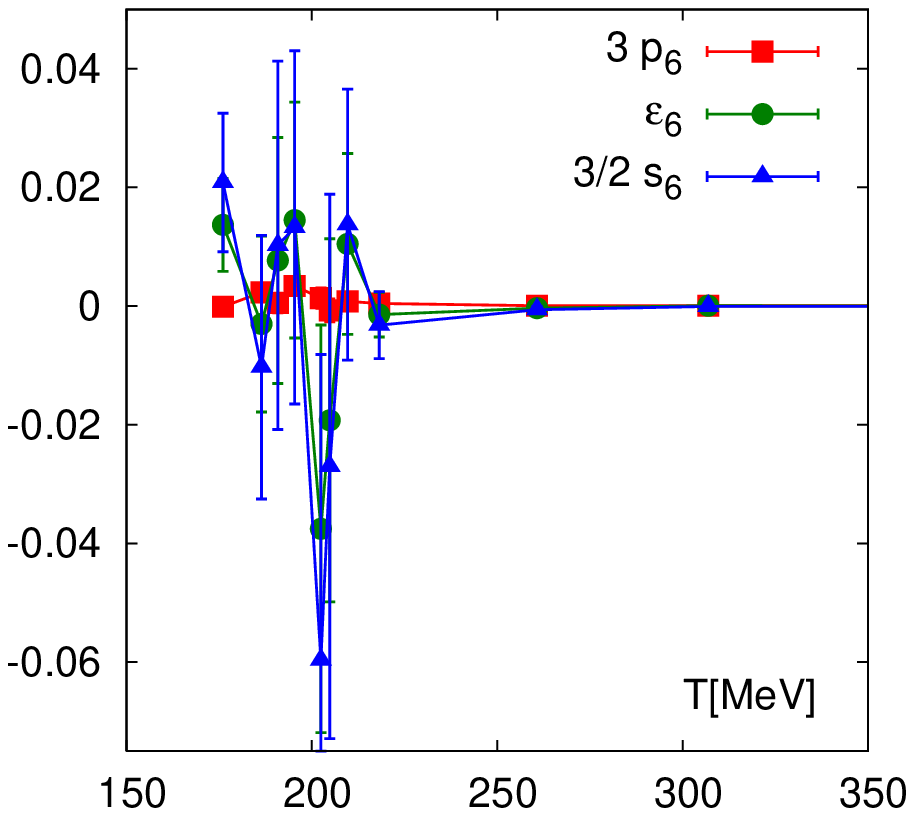}
}
\end{center}
\caption{Taylor coefficients of the pressure, the energy and entropy
density with respect to the baryon chemical potential. Results are
obtained with the p4fat3 action on $N_\tau=4$ (full) and $N_\tau=6$
(open symbols) lattices.}
\label{fig:coff}  
\end{figure}

Using standard thermodynamic relations we can also calculate the expansion 
coefficients of the trace anomaly or equivalently the 
difference between energy density and three times the pressure,
\begin{equation}
\frac{\epsilon - 3p}{T^4} = \sum_{n=0}^\infty \bar{c}_{n}^{B}(T) 
\left(\frac{\mu_B}{T}\right)^B,
\label{eq:e3pTaylor}
\end{equation}
where the expansion coefficients $\bar{c}_{n}^{B}$ are given by
\begin{equation}
\bar{c}_{n}^{B}(T) 
=T \frac{ {\rm d} c_{n}^{B} (T) }{{\rm d} T}.
\label{eq:cne3p}
\end{equation}
Combining Eqs.~\ref{eq:PTaylor_hadronic},~\ref{eq:e3pTaylor}, and 
\ref{eq:cne3p} we then obtain the Taylor expansions for the energy and 
entropy densities \cite{isen_eos}
\begin{eqnarray}
\frac{\epsilon}{T^4} &=& \sum_{n=0}^\infty \left(3 c_n^B(T) +
\bar{c}_n^B(T)\right) \left(\frac{\mu_B}{T}\right)^n 
              \equiv \sum_{n=0}^{\infty} \epsilon_n\left(\frac{\mu_B}{T}\right)^n,\\
\frac{s}{T^3} &\equiv& \frac{\epsilon +p-\mu_B n_B}{T^4}
= \sum_{n=0}^\infty \left( (4-n) c_n^B(T) +
\bar{c}_n^B(T)\right) \left(\frac{\mu_B}{T}\right)^n 
              \equiv \sum_{n=0}^{\infty} s_n\left(\frac{\mu_B}{T}\right)^n.
\label{eq:es}
\end{eqnarray}
At present, we calculate the expansion coefficients $\bar{c}_n^B$ from
the coefficients $c_n^B$, in accordance with Eq.~\ref{eq:cne3p}, by
performing the $T$ derivative numerically, which introduces a small
systematic error.

In Fig.~\ref{fig:coff} 
we show the second, fourth and sixth order expansion
coefficients of the pressure, energy density and entropy density as given
in Eqs.~\ref{eq:PTaylor_hadronic} and \ref{eq:es}, obtained with the
p4fat3 action. Full symbols are from $N_\tau=4$ lattices, while
the open symbols denote results from $N_\tau=6$ lattices. We again find 
small cut-off effects, however, higher order derivatives of pressure, energy
density and entropy density with respect to $\mu_B$ are still very 
preliminary, as the error bars are large. This is especially true for 
the results from $N_\tau=6$ lattices. Nevertheless, the overall pattern of the
coefficients is in agreement with expectations based on an analysis of the singular
behavior of the free energy, making use of an appropriate scaling Ansatz.

We find that the magnitude of the coefficients is decreasing drastically
with increasing order, for all analyzed temperatures. Thus an
approximation of the equation of state for small baryon chemical
potential by means of a fourth or sixth order expansion seems to be
justified. In general, an analysis of the radius of convergence of
such a Taylor series is of great interest for an analysis of the QCD
phase diagram, since the radius of convergence is bounded by the
location of the QCD critical point as well as by any first order phase
transition line (see Sec.~\ref{sec:radius}).

\section{The isentropic equation of state}
\label{sec:isen}
By using the Taylor expansion coefficients of the baryon number
(Eq.~\ref{eq:nB_hadronic}) and entropy density (Eq.~\ref{eq:es}), we
can compute the ratio of entropy per baryon number as function of $T$
and $\mu_B$. The 0-th order coefficient for the entropy density has been 
taken from \cite{EoS}. Solving numerically for a constant ratio of entropy per
baryon number, $s/n_B$, we determine isentropic trajectories in the
$(T,\mu_B)$-plane.  These trajectories are relevant for the
description of matter created in relativistic heavy ion collisions.
After equilibration the dense medium created in such a collision will
expand along lines of constant entropy per baryon. It then is of
interest to calculate thermodynamic quantities along such isentropic
lines.

\begin{figure}
\begin{center}
\resizebox{0.49\textwidth}{!}{%
  \includegraphics{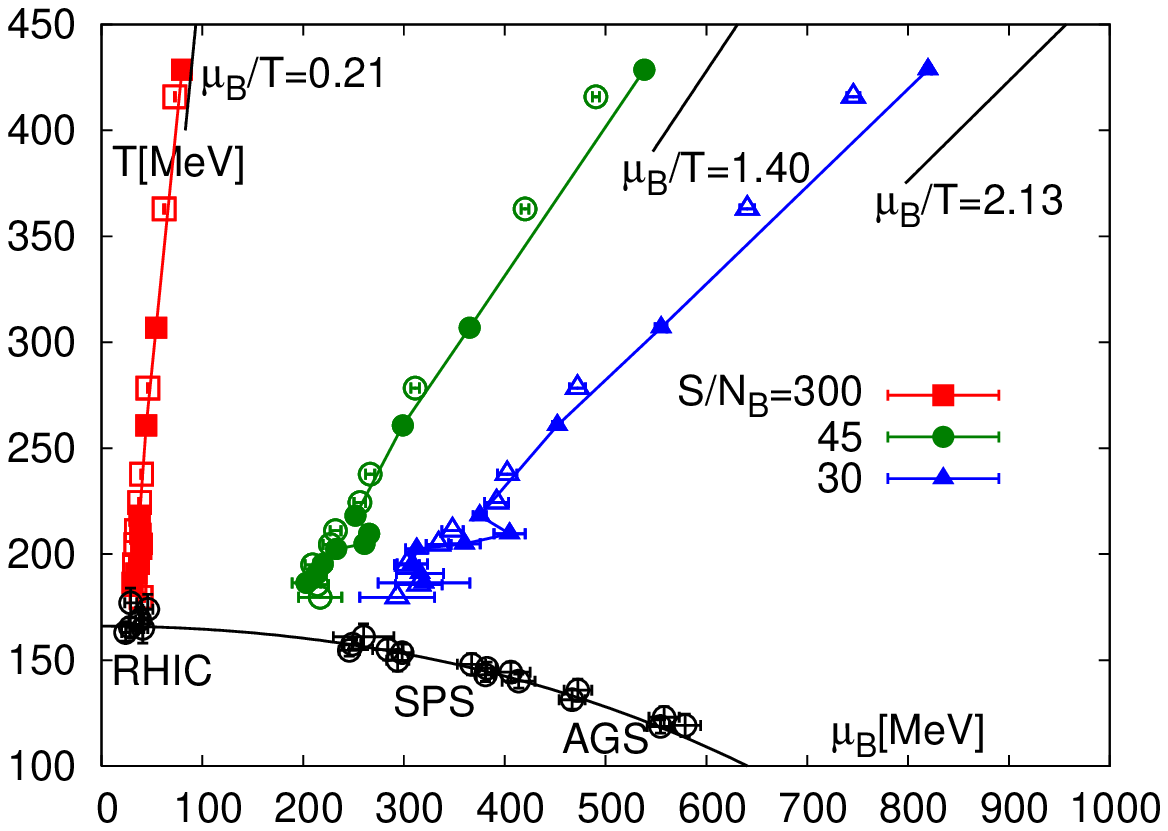}
}
\resizebox{0.49\textwidth}{!}{%
  \includegraphics{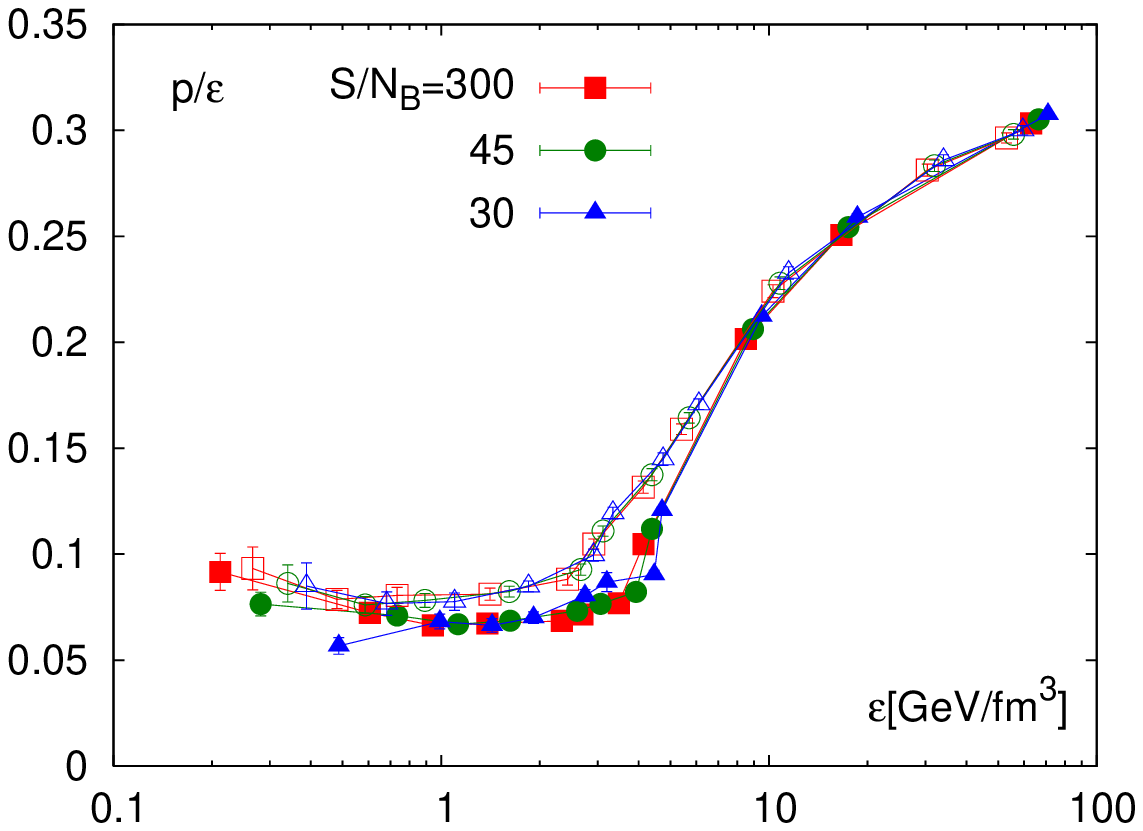}
}
\end{center}
\caption{On the left panel we show Isentropic trajectories in the
($T-\mu_B$)-diagram, corresponding to $s/n_B=300,45,30$ respectively.
Open symbols are form $N_\tau=6$, obtained by a 4th order Taylor
expansion of the pressure. Filled symbols are from $N_\tau=4$
calculations. We also show freeze-out data, as well as a
parameterization of the freeze-out curve from
\cite{redlich05}. Corresponding free gas limits are $\mu_B/T=0.21,
1.40, 2.13$ respectively and are indicated by solid lines.  On the
right panel we plot the the ratio of pressure and energy density along
those trajectories.}
\label{fig:isen_eos}
\end{figure}

We find that isentropic expansion at high temperature is well
represented by lines of constant $\mu_B/T$ down to temperatures close
to the transition, $T\simeq 1.2 T_0$. In the low temperature regime we
observe a bending of the isentropic lines in accordance with the
expected asymptotic low temperature behavior. The isentropic expansion
lines for matter created at SPS correspond to $s/n_B \simeq 45$ while
the isentropes at RHIC correspond to $s/n_B \simeq 300$. The energy
range of the AGS which also corresponds to an energy range relevant
for future experiments at FAIR/Darmstadt is well described by $s/n_B
\simeq 30$. These lines are shown in Fig.~\ref{fig:isen_eos} (left)
together with data points characterizing the chemical freeze-out of hadrons
measured at AGS, SPS and RHIC energies.  These data points have been
obtained by comparing experimental results for yields of various
hadron species with hadron abundances in a resonance gas
\cite{cleymans,redlich05}. The solid curve shows a phenomenological
parameterization of these {\it freeze-out data} \cite{redlich05}.  In
general our findings for lines of constant $s/n_B$ are in good
agreement with phenomenological model calculations that are based on
combinations of ideal gas and resonance gas equations of state at high
and low temperature, respectively \cite{shuryak,toneev}.

Results shown in Fig.~\ref{fig:isen_eos} are based on a fourth order
expansion of the pressure. We find, however, that the truncation
error is small, i.e. the results change only little when we
consider also the sixth order term in $\mu_B$. In accordance with the 
good convergence of our results, we find, that all trajectories shown 
in Fig.~\ref{fig:isen_eos} (left)
are well within the radius of convergence of the Taylor series.
At present we estimate the radius of convergence of the pressure series
to $(\mu_B/T)^{\rm crit}\gsim 2.7$. The cut-off effects can be
estimated by comparing open and full symbols.

We now proceed and calculate energy density and pressure on lines of
constant entropy per baryon number using our Taylor expansion results
up to ${\cal O}(\mu_B^4)$. Again we take the 0-th order coefficients from
\cite{EoS}. We find that both quantities obtain
corrections of about 10$\%$ at AGS (FAIR) energies ($s/n_B=30$) and
high temperatures.  The dependence of $\epsilon$ and $p$ on $s/n_B$
cancels to a large extent in the ratio $p/\epsilon$, which is most
relevant for the analysis of the hydrodynamic expansion of dense
matter. This may be seen by considering the leading ${\cal
O}(\mu_B^2)$ correction,
\begin{equation} \frac{p}{\epsilon} = \frac{1}{3} - \frac{1}{3}
\frac{\epsilon_0-3p_0}{\epsilon_0} \left( 1 + \left[
\frac{\bar{c}_2}{\epsilon_0-3p_0} - \frac{\epsilon_2}{\epsilon_0}
\right] \left( \frac{\mu_B}{T} \right)^2 \right) \; .
\label{povere}
\end{equation} In Fig.~\ref{fig:isen_eos} (right) we show $p/\epsilon$
as function energy density along our three isentropic trajectories. The
softest point of the equation of state is found to be
$(p/\epsilon)_{min} \simeq 0.07-0.09$,
for $N_\tau=4$ and $6$ respectively. Within our current numerical
accuracy it is independent of $s/n_B$. Similar results for the asqtad
action have been obtained in~\cite{milc_dens}. However, as our data is
preliminary, the analysis clearly suffers from insufficient statistics, which is 
in particular true for our $N_\tau=6$ results.

\section{Hadronic Fluctuations}
Quark number fluctuations are related to the derivatives of QCD
partition function with respect to the quark chemical potentials by the
fluctuation-dissipation theorem. The Taylor expansion coefficients
$c_{i,j,k}^{u,d,s}$, as defined in Eq.~\ref{eq:PTaylor}, can thus be
directly interpreted as quark number fluctuations at $\mu=0$.
However, quark fluctuations can not be detected directly in
experiments due to confinement. Therefore we will consider
fluctuations in terms of hadronic quantum numbers, i.e. baryon number
$B$, electric charge $Q$ and strangeness $S$, which are more easily
obtained by experiment. These fluctuations are related with the
hardonic Taylor expansion coefficients $c_{i,j,k}^{B,Q,S}$, as given in
Eq.~\ref{eq:PTaylor_hadronic}.

In general, the quadratic fluctuations $\chi_2^X$ at zero chemical
potentials can be obtained from the second order coefficient $c_2^X$
\begin{equation}
  c_2^X \equiv \left.\frac{1}{2VT^3}
    \frac{\partial^2 \ln Z}{\partial(\mu_X/T)^2}\right|_{\muBQS=0}
    = \frac{1}{2VT^3} \langle(\delta N_X)^2\rangle_0\ ,
\end{equation}
where $\delta N\equiv N-\langle N\rangle$ denotes the normalized net-density and
$\langle\ldots\rangle_0$ indicates that the expectation value has been
taken at $\muBQS=0$. Under such conditions, baryon number, electric
charge and strangeness vanish, and we have $\delta N=N$. We define 
\begin{equation}
  \chi_2^X = \frac{1}{VT^3} \langle N_X^2\rangle_0 = 2c_2^X
\end{equation}
Similarly, we obtain the quartic charge fluctuations by
\begin{equation}
  \chi_4^X = \frac{1}{VT^3} \left( \langle N_X^4 \rangle_0
    - 3 \langle N_X^2 \rangle_0^2 \right) = 24c_4^X\ ,
\end{equation}
and correlations among two conserved charges by
\begin{eqnarray}
   \chi_{11}^{XY} &=& \frac{1}{VT^3} \left( \langle N_XN_Y \rangle_0 
     - \langle N_X \rangle_0 \langle N_Y \rangle_0 \right)
   	= c_{11}^{XY}\ ,\\
   \chi_{22}^{XY} &=& \frac{1}{VT^3} \left( \langle N_X^2N_Y^2 \rangle_0 
     - \langle N_X^2\rangle_0\langle N_Y^2\rangle_0 
   	-2\langle N_XN_Y\rangle_0^2 \right) = 4c_{22}^{XY},
\end{eqnarray}
where $X,Y \in \{B,Q,S\}$.

\begin{figure}
\begin{center}
\resizebox{0.49\textwidth}{!}{%
  \includegraphics{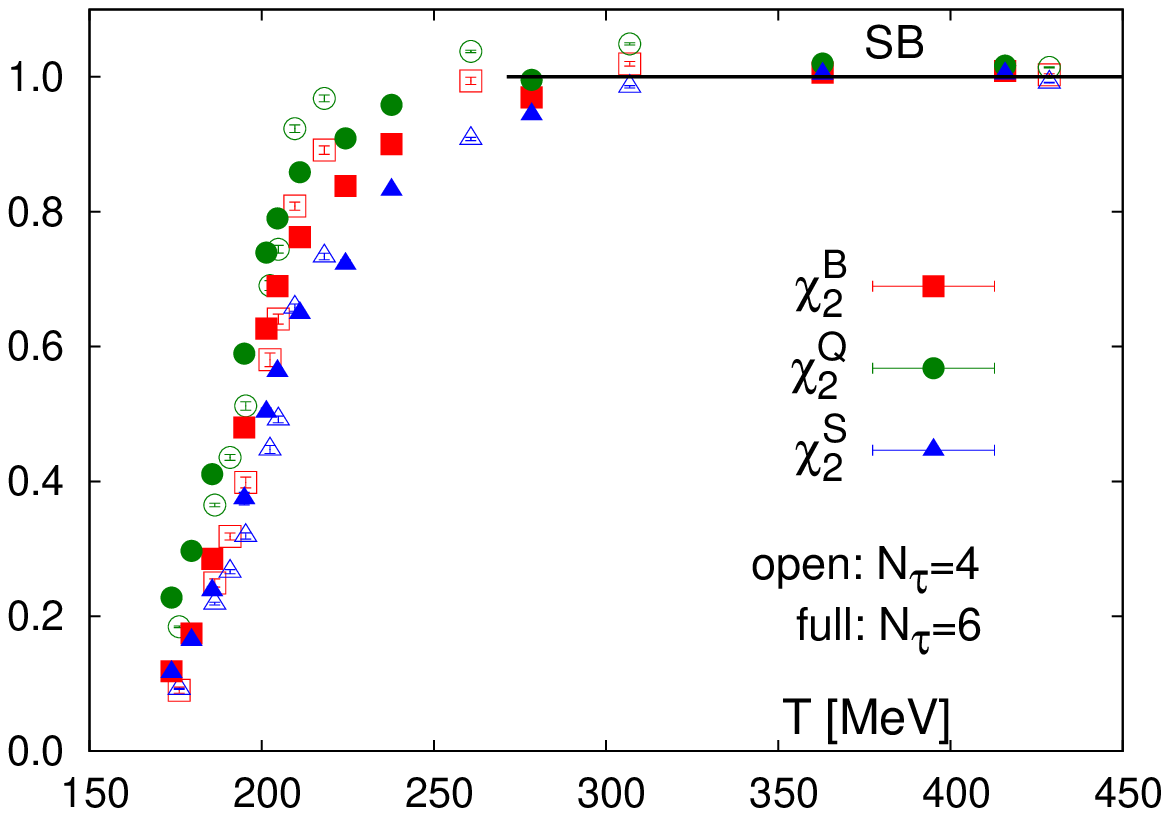}
}
\resizebox{0.49\textwidth}{!}{%
  \includegraphics{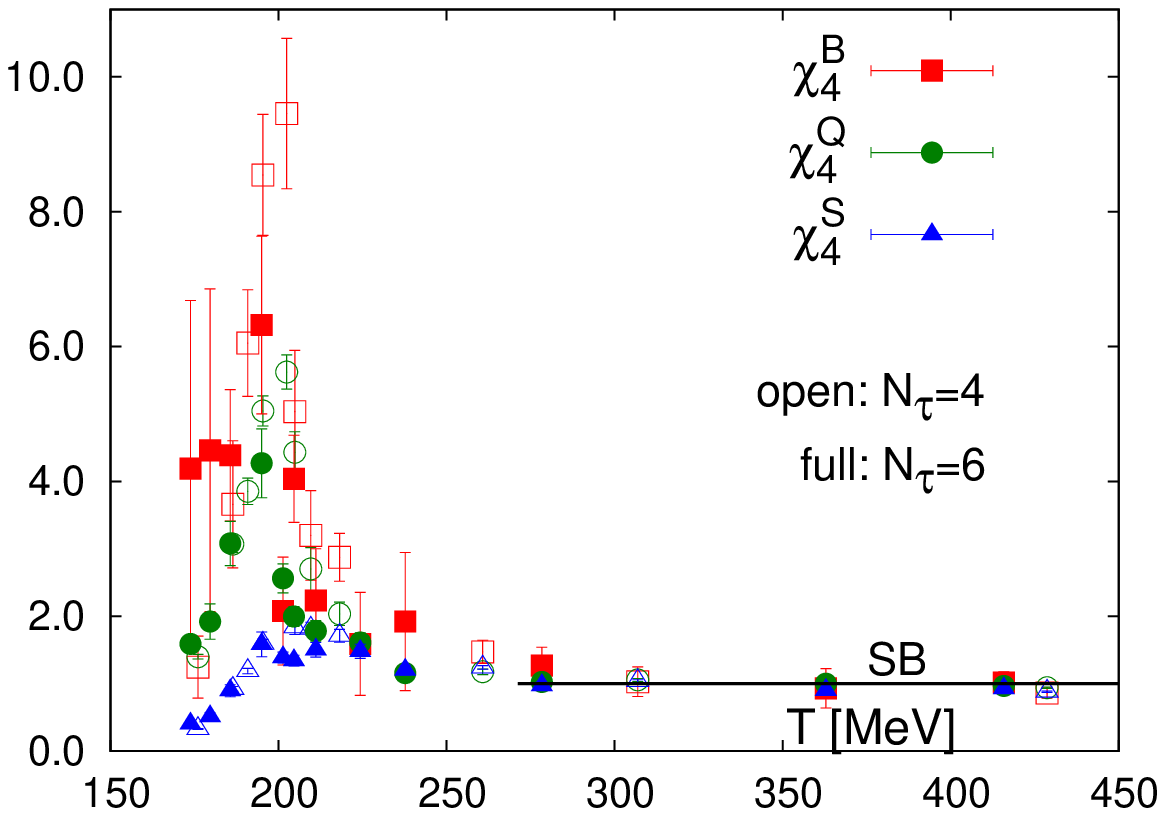}
}
\end{center}
\caption{
Quadratic and quartic fluctuations of baryon number, electric charge
and strangeness, normalized by their corresponding Stefan-Boltzmann value.
The results on $N_\tau=4$ lattices (open symbols)
and $N_\tau=6$ lattices (full symbols) are in good agreements.
}
\label{fig:c24BQS}  
\end{figure}

In Fig.~\ref{fig:c24BQS} we show results for quadratic and quartic 
fluctuations of $B$, $Q$ and $S$.  The quadratic fluctuations $\chi_2^{B,Q,S}$
rise rapidly in the transition region where the quartic fluctuations 
$\chi_4^{B,Q,S}$ show a peak. The peak height is more pronounced for the baryon
number fluctuations than for fluctuations of the strange quarks. 

We compare the results obtained on lattices with temporal extent $N_\tau = 4$
and $6$. We notice that they are in general compatible with each other, 
especially in the high temperature phase, where both quadratic and quartic
fluctuations approach the Stephan-Boltzmann limit quickly. The transition
temperature has been previously determined to be $T_c = 202$ MeV and $196$ MeV
on $N_t = 4$ and $6$ lattices respectively \cite{Tc}. We thus conclude that at
temperatures of about
$1.5T_c$ and higher, quadratic and quartic fluctuations of $B$, $Q$ and $S$
are well described by the ideal massless quark gas.

At low temperature, hadrons are the relevant degrees of freedom. 
The hadron resonance gas (HRG) model has been shown to provide a 
good description of thermal conditions at freeze-out. We thus compare
the fluctuations in the low temperature phase with a HRG model, where
we include all mesons and baryons with masses smaller than 2.5 GeV from 
the particle data book. 

\begin{figure}
\begin{center}
	\resizebox{0.49\textwidth}{!}{\includegraphics{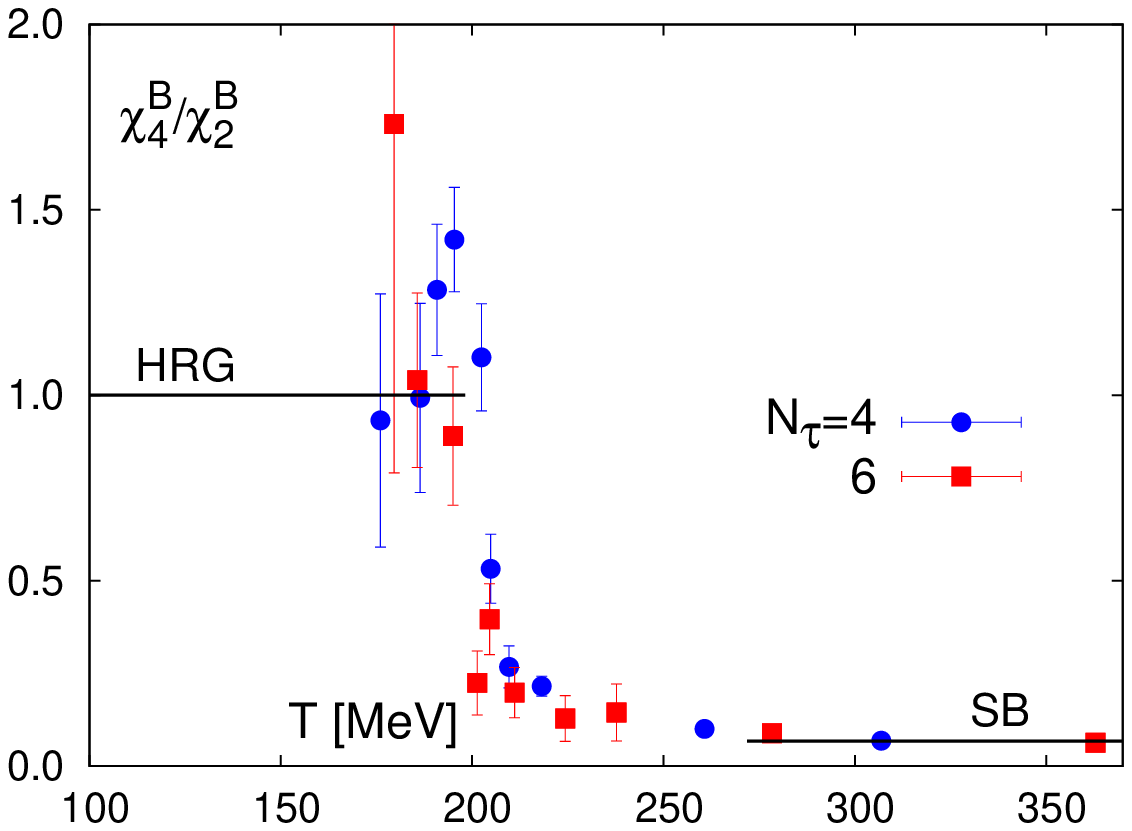}}
	\resizebox{0.49\textwidth}{!}{\includegraphics{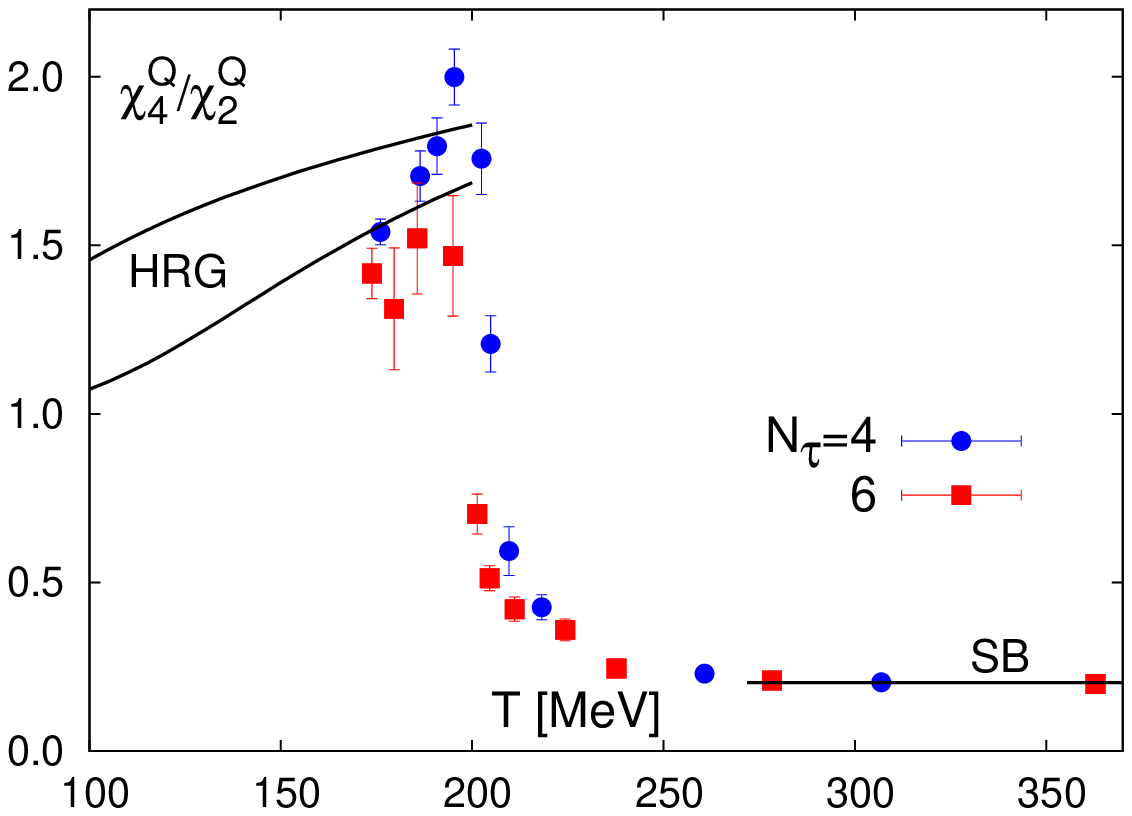}}
	\resizebox{0.49\textwidth}{!}{\includegraphics{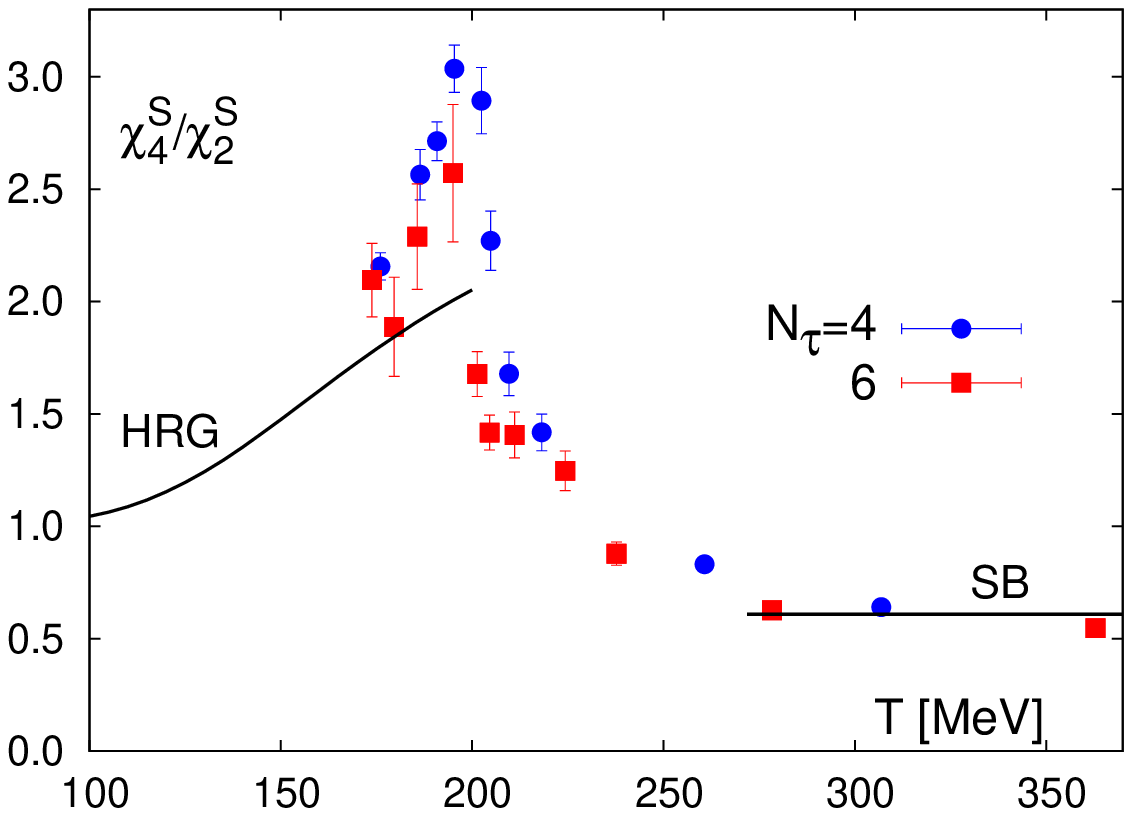}}
\end{center}
\caption{
Ratio of quartic and quadratic fluctuations of baryon number ($B$), strangeness ($S$)
and electric charge ($Q$). The two curves of the HRG model in the top right
figure correspond to charge fluctuations with physical pions (upper) and
infinitely heavy pions (lower curve).
}
\label{fig:R2BQS}
\end{figure}

In Fig.~\ref{fig:R2BQS}, we show the ratio of quartic and quadratic
fluctuations for $B$, $S$ and $Q$.  
In the HRG model, $\chi_4^B/\chi_2^B$ is 
easily obtained in the Boltzmann approximation, which is valid for 
a dilute baryonic gas in the temperature range of interest. One finds
that all details on the hadron mass spectrum and temperature dependence 
cancel and the result is a constant, given by the unit square of the 
baryonic charge (one for all baryons). This is
in fact reproduced by the lattice results shown in Fig.~\ref{fig:R2BQS} 
(top left).

The ratio of quartic and quadratic fluctuations
for $S$ and $Q$ are more complicated even in the
Boltzmann limit, since hadrons with different electric/strange charge
give rise to different contributions to the corresponding fluctuations. For
strangeness fluctuations, shown in Fig.~\ref{fig:R2BQS} (bottom), the Boltzmann
limit is still a good approximation; but for electric charge fluctuations,
the pion mass plays an important role. In order to check for the sensitivity
of electric charge fluctuations on the pion mass, we show in Fig.~\ref{fig:R2BQS}
(top right)
results of a HRG model calculation with physical pion masses and without
the pion sector, i.e. for infinitely heavy pions.

\begin{figure}
\begin{center}
	\resizebox{0.49\textwidth}{!}{\includegraphics{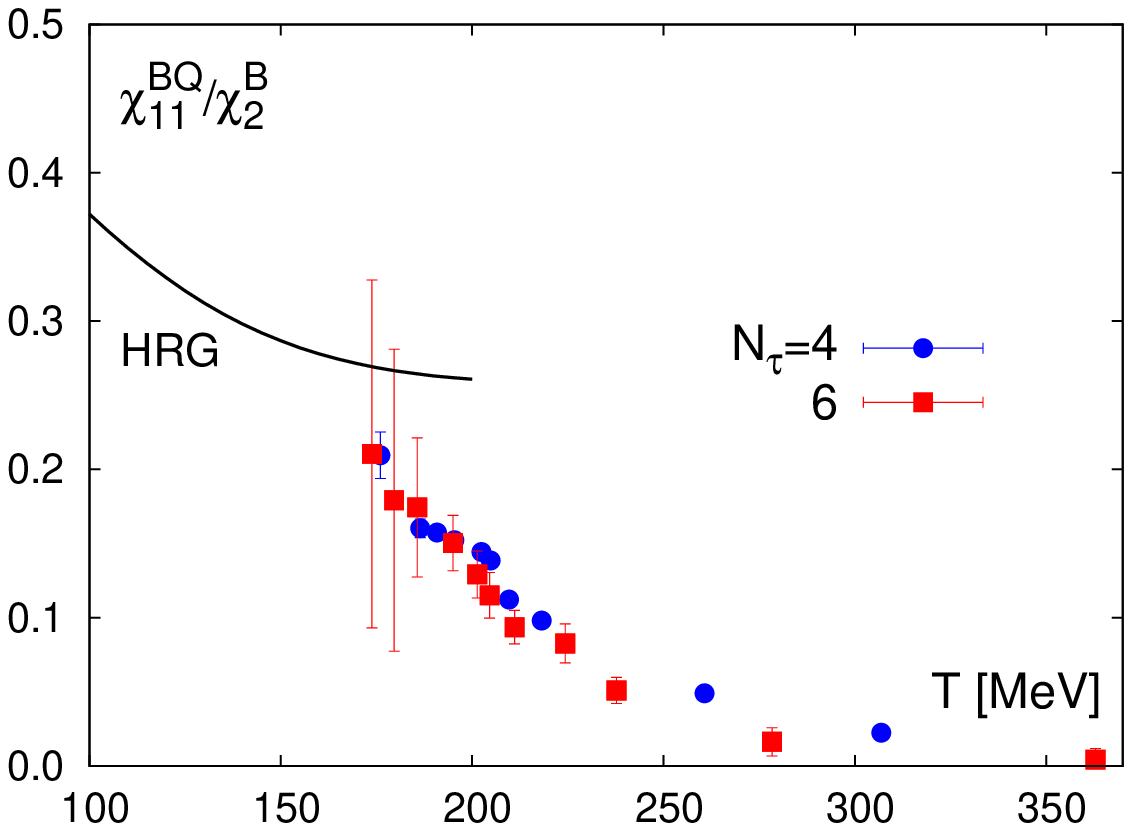}}
	\resizebox{0.49\textwidth}{!}{\includegraphics{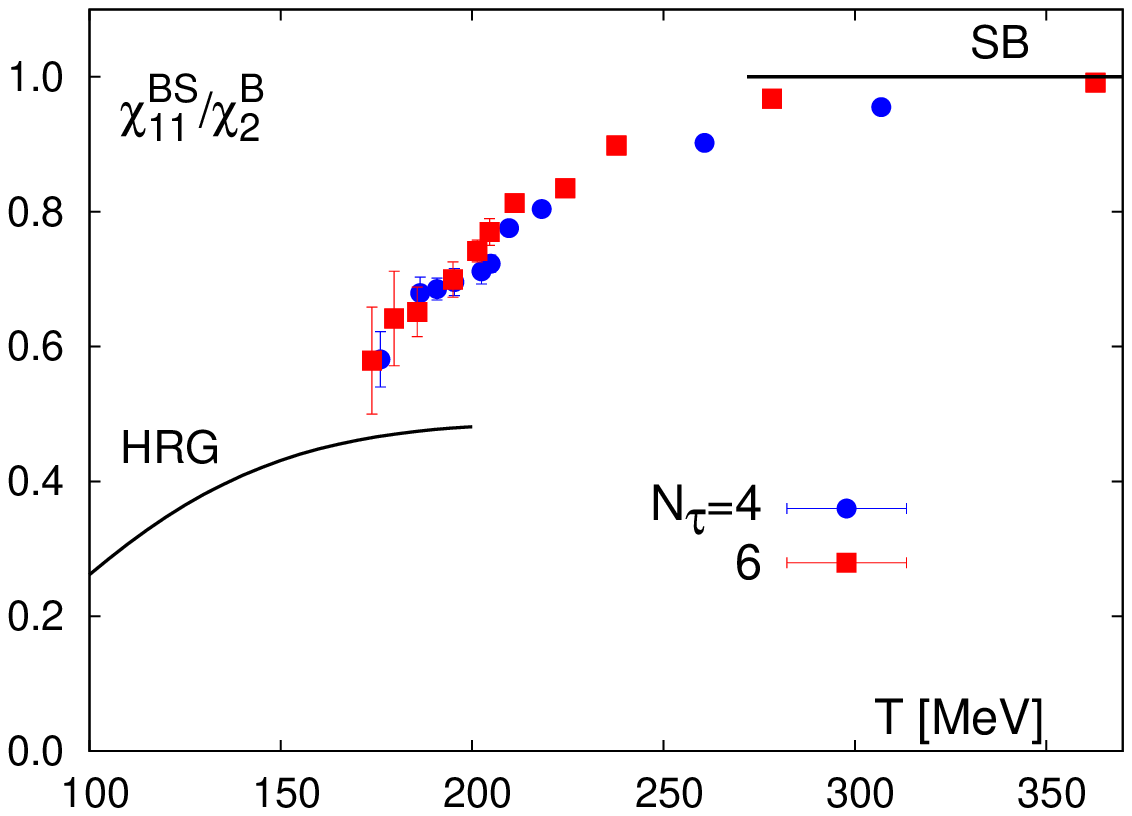}}
	\resizebox{0.49\textwidth}{!}{\includegraphics{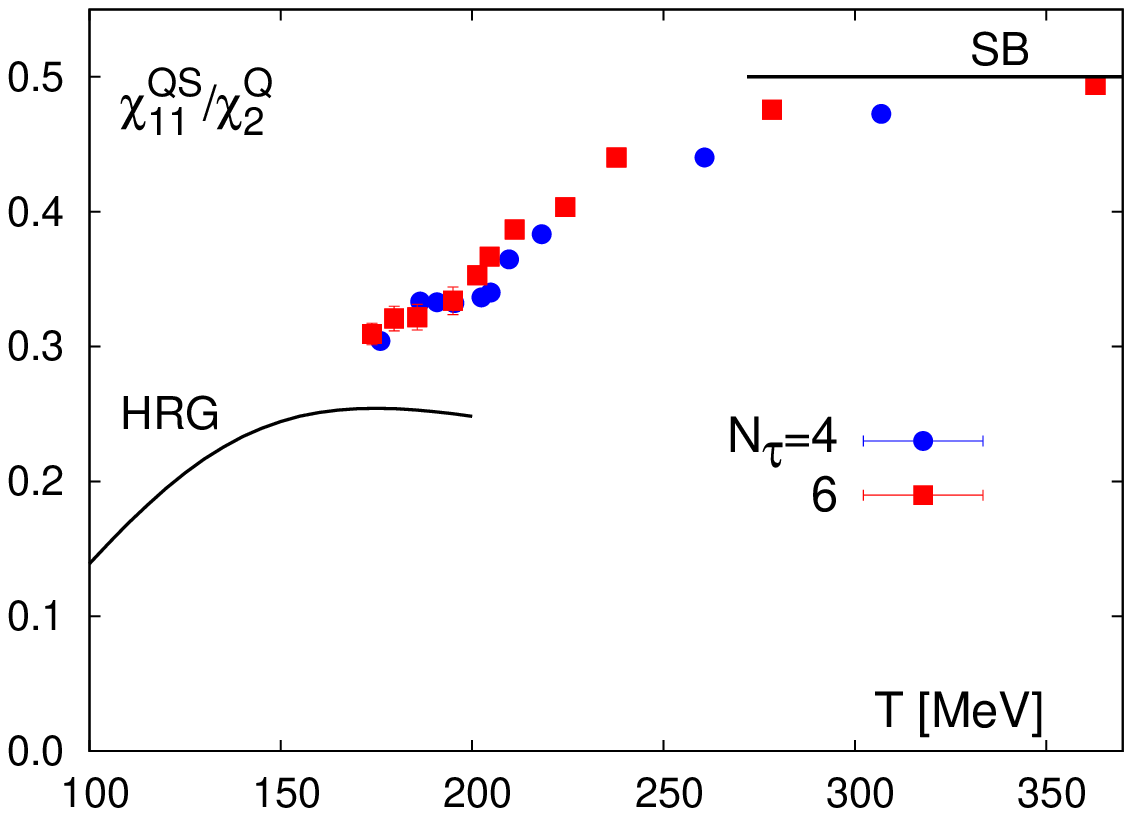}}
\end{center}
\caption{
Pairwise correlatios of the conserved charges baryon number (B),
electric charge (Q) and strangeness (S) as function of the
temperature, normalized to the quadratic fluctuations of B and Q
respectively.
}
\label{fig:R11BQS}
\end{figure}

In Fig.~\ref{fig:R11BQS}, we show the various correlations $\chi_{11}^{BQ}$,
$\chi_{11}^{BS}$ and 
$\chi_{11}^{QS}$ normalized to quadratic fluctuations $\chi_2^B$ and $\chi_2^Q$
respectively. The results from $N_\tau=4$ and $6$ lattices agree with each other
very well, and they are compared with the HRG model in the low temperature phase
and Stephan-Boltzmann limit in the high temperature phase.
We find that the correlations reproduce the qualitative behaviour of the HRG
model below $T_c$ and again start to agree with the free gas predictions
for $T\gsim 1.5T_c$. 

\section{Radius of convergence}
\label{sec:radius}
Whenever an observable at $\mu_B> 0$ is approximated by a truncated
Taylor series, expanded around $\mu_B=0$, an analysis of the
convergence behavior of the series is, of course, mandatory.
In addition, the convergence radius of such a series, as {\it e.g.}
given in Eq.~\ref{eq:PTaylor_hadronic} for the pressure, is of great
interest for the structure of the QCD phase diagram. By definition,
the radius of convergence is the distance to the closest singularity in
the $(T,\mu_B)$-plane. It is bounded by the QCD critical point as well
as by any line of first order transitions and could thus
provide an interesting method to determine the QCD critical point. This
method was proposed in Ref.~\cite{details1} and was applied in the
case of standard staggered fermions in Ref.~\cite{GG}.

One way to evaluate the radius of convergence $\rho$ is by means of
the formula
\begin{equation}
\rho(T)=\lim_{n\to\infty}\rho_n(T),\qquad\mbox{with}\qquad 
\rho_n(T)=\sqrt{{c_n^B(T)}/{c_{n+2}^B(T)}}.
\label{eq:rho}
\end{equation}
In practice, we can not perform the limit $n\to\infty$, instead we consider the
first few approximations $\rho_n(T)$, with $n\lsim 6$.
The second approximation for $\rho(T)$ is given
by
$\rho_2(T)\equiv\sqrt{c_2^B(T)/c_4^B(T)}=\sqrt{12}/\sqrt{\chi_4^B/\chi_2^B}$.
We see that $\chi_4^B/\chi_2^B$ as shown in
Fig.~\ref{fig:R2BQS} is closely related to the radius of
convergence. If this quantity develops a peak,
which increases above the resonance gas value of
${\chi_4^B/\chi_2^B}=1$, one might expect the critical point to be
found at $(\mu_B/T)^{\rm crit}\lsim \sqrt{12}$.  Of course, 
higher order approximations for the radius of convergence are needed 
to substantiate such an estimate and to establish the existence of a
critical point.

Our results are at present not conclusive, as the
$N_\tau=6$ data shown in Fig.~\ref{fig:R2BQS}(left) suffers from
insufficient statistics and temperature resolution.  Note, however, that
the peak in the baryon number fluctuation ratio $\chi_4^B/\chi_2^B$
will develop in the chiral
limit as can be seen from a scaling analysis of the free energy and
has recently also shown in a chiral model calculation \cite{redlich}.

\section{Conclusions}
\label{sec:con}

We have calculated corrections to the equation of state
arising from a non-zero baryon chemical potential, by means of a
Taylor expansion of the pressure. Within this framework we calculated
the isentropic equation of state along lines of constant entropy per
baryon number ($s/n_B$) for RHIC, SPS and AGS (FAIR) energies. Within
our current, preliminary, analysis we find the softest point of the
equation of state to be independent of $s/n_B$.

Furthermore, we have analyzed the quadratic and quartic fluctuations
of baryon number, electric charge and strangeness, as well  their
ratios. We find these quantities to be in good agreement
with the free gas results at temperatures of $T>1.5T_c$. Below $T_c$, 
qualitative features of the resonance gas are reproduced. This ratio
for the baryon number is closely related to the second approximation of 
the convergence radius of the Taylor series of the pressure with respect
to the baryon chemical potential.

\section*{Acknowledgments}
We thank all members of the RBC-Bielefeld Collaboration for helpful
discussions and comments.
This work has been supported in part by contracts DE-AC02-98CH10886
and DE-FG02-92ER40699 with the U.S. Department of Energy.
Numerical simulations have been performed on the QCDOC super-computer of 
the RIKEN-BNL research center, the DOE
funded QCDOC at Brookhaven National Laboratory (BNL) and the apeNEXT 
installation at the University of Bielefeld.


\begin{thebibliography}{99}
\bibitem{overview}
M.P.~Lombardo, J. Phys. G \textbf{35} (2008) 104019; \\
C. Schmidt, PoS \textbf{LAT2006} (2006) 021.
\bibitem{eos6}
C.~R.~Allton, M.~Doring, S.~Ejiri, S.J.~Hands, O.~Kaczmarek,
F.~Karsch, E.~Laermann, K.~Redlich, Phys. Rev D \textbf{71} (2005) 054508.
\bibitem{details1}
C.~R.~Allton \textit{et al.}, Phys. Rev. D \textbf{66} (2002) 074507.
\bibitem{details2}
C.~Miao and C.~Schmidt, PoS \textbf{LAT2007} (2007) 175.
\bibitem{EoS}
M. Cheng \textit{et al.}, Phys. Rev. D \textbf{77} (2008) 014511.
\bibitem{p4fat3}
U.~M.~Heller, F.~Karsch and B.~Sturm, Phys. Rev.  D \textbf{60} (1999) 114502.
\bibitem{Hegde}
P.~Hegde, F.~Karsch, E.~Laermann and S.~Shcheredin,
Eur. Phys. J.  C {\bf 55} (2008) 423.
\bibitem{milc_dens}
C.~Bernard \textit{et al.}, Phys. Rev.  D \textbf{77} (2008) 014503.
\bibitem{GG}
R.~V.~Gavai and S.~Gupta, Phys. Rev.  D \textbf{71} (2005) 114014;
R.~V.~Gavai and S.~Gupta, arXiv:0806.2233 [hep-lat]; S.~Gupta this proceedings.
\bibitem{isen_eos}
S.~Ejiri, F.~Karsch, E.~Laermann and C.~Schmidt,
Phys. Rev. D \textbf{73} (2006) 054506.
\bibitem{cleymans}
J.~Cleymans and K.~Redlich, Phys. Rev. Lett. \textbf{81} (1998) 5284;\\
J.~Cleymans and K.~Redlich,  Phys. Rev. C \textbf{60} (1999) 054908.
\bibitem{redlich05}
J.~Cleymans, H.~Oeschler and K.~Redlich, S. Wheaton, Phys. Rev. C \textbf{73} (2006) 034905.
\bibitem{shuryak}
C. M. Hung and E. Shuryak, Phys. Rev. D \textbf{57} (1998) 1891.
\bibitem{toneev}
V.~D.~Toneev, J.~Cleymans, E.~G.~Nikonov, K.~Redlich and A.~A.~Shanenko,
J. Phys. G \textbf{27} (2001) 827.
\bibitem{Tc}
M.~Cheng {\it et al.}, Phys. Rev. D \textbf{74} (2006) 054507.
\bibitem{redlich}
B.~Stokic, B.~Friman and K.~Redlich, arXiv:0809.3129 [hep-ph].
\end{thebibliography}
\end{document}